\documentclass[preprint,12pt,a4paper]{elsarticle_custom}
\usepackage{graphicx} 
\usepackage{epstopdf}
\usepackage{tikz}
\usetikzlibrary{positioning}
\usetikzlibrary{arrows}
\usetikzlibrary{calc}
\usepackage{siunitx}
\usepackage{amsmath,amssymb,amsfonts}
\usepackage[utf8]{inputenc}
\usepackage[T1]{fontenc}
\usepackage[figuresright]{rotating}
\usepackage{hyperref}
\usepackage{tikz-cd}

\DeclareGraphicsExtensions{.eps}
\newcommand{\euler}{\mathrm{e}}
\newcommand{\figref}{Fig.~\ref}

\newtheorem{proposition}{Proposition}
\newtheorem{definition}{Definition}[section]
\newtheorem{remark}{Remark}

\begin{document}

\begin{frontmatter}

\tnotetext[footnoteinfo]{
This work was supported by the Czech Science Foundation project No. 21-07321S, and by the European Union under the project Robotics and advanced industrial production No. $CZ.02.01.01/00/22\_008/0004590$. The work of the first and last authors was also supported by a public grant overseen by the French National Research Agency (ANR) as part of the « Investissements d’Avenir » program, through the "ADI 2020" project funded by the IDEX Paris-Saclay, ANR-11-IDEX-0003-02. The work of the first author was also supported by the Grant Agency of the Czech Technical University in Prague , student grant No. SGS23/157/OHK2/3T/12.\\ \# Corresponding author: {\tt tomas.vyhlidal@fs.cvut.cz} \\ \\
\textbf{Cite as:}\space Can Kutlu Yüksel, Tomáš Vyhlídal, Jaroslav Bušek, Martin Hromčík, Silviu-Iulian Niculescu. \textit{A spectrum-based filter design for periodic control of systems with time delay.} Journal of Sound and Vibration, Volume 604, 2025, 118959. https://doi.org/10.1016/j.jsv.2025.118959 \\ \\
\copyright 2025 The Authors. Published by Elsevier Ltd. This is an open access article under the CC BY license (\href{https://creativecommons.org/licenses/by/4.0/}{https://creativecommons.org/licenses/by/4.0/}).} 

\title{A Spectrum-based Filter Design for Periodic Control of Systems with Time Delay$^*$}

\author[inst1,inst2]{Can Kutlu Y{\"u}ksel}
\author[inst1]{Tom\'{a}\v{s} Vyhl\'{i}dal $^{\#}$}
\author[inst1]{Jaroslav Bu\v{s}ek}
\author[Martin]{Martin Hrom\v{c}\'{i}k}
\author[inst2]{Silviu-Iulian Niculescu}

\address[inst1]{Department of Instrumentation and Control Engineering, Faculty of Mechanical Engineering, Czech Technical University in Prague, Technická 4, Prague 6, 16607, Czechia.}
\address[inst2]{Universit\'{e} Paris-Saclay,
        CNRS, CentraleSup\'{e}lec, Inria, Laboratoire des Signaux et Syst\`{e}mes, 91192
        Gif-sur-Yvette, France}
\address[Martin]{Department of Control Engineering, Faculty of Electrical Engineering, Czech Technical University in Prague, Karlovo n\'{a}m\v{e}st\'{i} 13,
121 35 Prague 2, Czechia}

\begin{abstract}
A fully analytical controller design is proposed 
to tackle a periodic control problem for stable linear systems with an input delay. Applying the internal model control scheme, the controller design reduces to designing a filter, which is done through the placement of poles and zeros. The zeros are placed to compensate for the harmonics and to achieve the desired degree of properness for the filter. For placing the poles, a quasi-optimal procedure is proposed utilizing the standard LQR method. Given the high-dimensionality of the filter due to targeting a large number of harmonics, the design, as well as controller implementation, is performed over a state-space representation.   
A thorough experimental case study is included to demonstrate both the practical feasibility and effectiveness of the proposed control design.  The experimental validation is performed on a physical system, the goal of which is to reject periodic vibrations acting on a mass-spring-damper setup where the sensor and the actuator are non-collocated.
\end{abstract}


\end{frontmatter}

\section{Introduction}
Controlling systems to behave in a repetitive fashion has been a long-lasting demand. Such capability can help isolate the controlled object from periodic disturbances caused by internal or external sources such as vibrations. The celebrated \emph{Internal Model Principle} (IMP) \cite{francis1976internal} unravels what is required for the control system and, thus, gives a roadmap for building controllers that can make a system track or reject a signal $v(t)$ with a certain period $T$, i.e. $v(t)=v(t+T)$. The design process begins by choosing a sufficiently accurate model for the periodic signal that captures the desired or observed periodic signal. The accuracy of this periodic model is determined by the number of harmonics it takes into account, i.e. the components in its Fourier series expansion given by
\begin{equation}
    v(t) =\frac{c_0}{2}+\sum_{l=1}^\infty c_l\cos\left(\frac{2\pi l}{T}t-\varphi_l\right),
    \label{eq:Fourier}
\end{equation}
with weights $c_0, c_l$ and phase shifts $\varphi_l, l=1..\infty$. One can choose a finite-dimensional signal model by simply combining the harmonic frequencies that matter for the application as in 
\begin{equation}
    V(s) = \frac{1}{s} \prod_{i=1}^k \frac{1}{s^2+\omega_i^2},
    \label{eq:finite_signal_model}
\end{equation}
where $\omega_i$ are the finitely many targeted harmonic frequencies of the decomposed signal in \eqref{eq:Fourier}. On the other hand, by 
an appropriate choice of the delays in the design, one can alternatively form an infinite-dimensional signal model that can count for all harmonics as in
\begin{equation}
    V(s) = \frac{1}{1-\euler^{-sT}},
    \label{eq:infinite_signal_model}
\end{equation}
which ideally captures every periodic signal satisfying $v(t)=v(t+T)$. Once the signal model is chosen, the control design proceeds to finding a controller that is capable of stabilizing the ``overall'' closed-loop that comprises the system and the chosen signal model. More precisely, in this case, finding a stabilizing controller results in placing the poles of the signal model as the zeros of a stable sensitivity function, which in return ensures asymptotic tracking/rejection of the signal.

Naturally, with the motivation to get the best periodic performance, there have been attempts to design the controller stemming from IMP with an infinite-dimensional signal model in \eqref{eq:infinite_signal_model}, \cite{hara1988repetitive}, \cite{chang1995analysis}. Inspired by such ideas, a notable generalization of the signal model which utilizes more delays compared to \eqref{eq:infinite_signal_model} showed that the increase in the number of delays can help drastically improve transient performance and robustness against variations in the period \cite{inoue1990practical}, \cite{chang1995analysis}, \cite{pipeleers2008robust}. Such controller designs based on an infinite-dimensional signal model are commonly classified under \emph{repetitive control} and have been considered for time delay systems \cite{omura2015attenuation}, nonlinear systems \cite{verrelli2023adaptive}, \cite{astolfi2021repetitive}, \cite{reinders2023repetitive},  MIMO systems \cite{weiss1999repetitive}, \cite{mirkin2020dead} and distributed systems \cite{macchelli2018dissipativity}. Nevertheless, a closer look at these approaches reveals that the closed-loop in discrete time or the modifications required to make the closed-loop stabilizable in continuous time due to the facts given in \cite{partington2004h}, \cite{gumussoy2018feedback} result in eventually targeting only finitely many harmonics properly. This fact partially justifies the design approaches that start with a finite-dimensional signal model in contrast to the infinite-dimensional one \cite{hillerstrom1994repetitive}, \cite{astolfi2022harmonic}, \cite{bajodek2023comparison}.
However, regardless of the chosen method, stabilization of the closed-loop system remains challenging when the controlled system suffers from input/output time delays since, in such a case, the closed-loop corresponds to an infinite-dimensional system of at least retarded-type \cite{michiels2014stability}, \cite{michiels2003delay}. Nevertheless, as in the case mentioned above, introducing more delays as controller parameters can help better stabilize the feedback \cite{hu2004using}. 


In this paper, we propose a straightforward, systematic controller design that can achieve periodic control of systems that suffer from input/output time delays with the use of finite-dimensional signal models and adjustable time delays. In particular, we aim for systems approximated by stable models with input delays but with no non-minimum-phases zeros which widely represent industrial processes \cite{grimholt2012optimal} and exploit the features they bring. The overall controller design follows the footsteps of the previous works of the authors \cite{yuksel2023harmonic}, \cite{yuksel2023distributed}, which were limited to a specific class of systems made approximated by a first-order model and time delay thanks to additional inner control loops. Since stable systems are considered, the framework chosen for the control system is \emph{Internal Model Control} (IMC) \cite{garcia1982internal}. We note that the chosen control framework does not fall far apart from the alternative methods proposed in the literature since it can be extended to repetitive control when taken into account with Youla-Ku\v{c}era parameterization as in \cite{chen2013new}. The conditions required for tracking/rejecting a periodic signal and delay compensation are derived completely analytically using the IMC framework. Subsequently, stabilization and performance of the control system are ensured by placing poles manually and with respect to a quadratic cost using the novel interpretation of the controller as a control system on its own. Furthermore, the effectiveness of the overall methodology is confirmed experimentally.
%
Note that the first ideas, as well as some preliminary results, were presented in a conference publication \cite{yuksel2023spectrum}, where only the controller structure was proposed and its validation was performed by simulations. Besides, the pole locations problem remained unaddressed.

\section{Controller design}
The proposed control design treats a stable plant
\begin{equation}
    G_\tau(s) = \frac{Y(s)}{U(s)}=G(s)\euler^{-s\tau},
    \label{eq:Gtau}
\end{equation}
with single input $u$ and single output $y$ ($U(s)$, $Y(s)$ in the Laplace form), where $\tau\in\mathbb{R}$, $\tau>0$ is the delay and  
\begin{equation}
    G(s) = \frac{a_\alpha s^\alpha+a_{\alpha-1}s^{\alpha-1}+ ... + a_1s+a_0}{b_{\beta}s^\beta+b_{\beta-1}s^{\beta-1}+ ... + b_1s+b_0},
    \label{eq:G}
\end{equation} 
with the polynomials in the numerator and the denominator being Hurwitz and satisfying $\beta,\alpha\in\mathbb{N}$, $\beta \geq \alpha$, i.e., the delay free system part is stable, proper and with minimum phase characteristic. 

The control design objective is to fully compensate $k$ dominant harmonics of the periodic exogenous signal $v(t)$ of the form 
(\ref{eq:finite_signal_model}), acting as the system disturbance $d(t)$ ($D(s)$ in Laplace form). This will be done by adopting the IMC control feedback arrangement shown in \figref{fig:IMC_scheme}. For the sake of generality, we assume that the disturbance $d$ acts at the system output. Since the system is assumed to be stable, a disturbance acting at a different part of the closed-loop can easily be projected to the output disturbance $d(t)$. 

To motivate the assumptions made for the considered systems, note that the first assumption of system \eqref{eq:G} being stable is a pre-requisite of applying the IMC framework. The second assumption of system \eqref{eq:G} being minimum phase is in order to obtain a stable IMC controller that comprises an inverse model of the delay-free system dynamics, as discussed below.


\subsection{Internal Model Controller}
\begin{figure}[t]
    \centering
    \scalebox{1.2}{\begin{tikzpicture}[
tf/.style = {rectangle, draw=black, fill=white,  thick, minimum width=60pt, minimum height=30pt},
sum/.style = {circle,draw=black,fill=white, thick, minimum size=15pt}]

\node[tf,label={above:Controller}] (Q)   {$Q(s)\mathrm{e}^{-s\theta}$};
\node[tf,label={above:System}] (G)   [right=30pt of Q]{$G_s(s)\mathrm{e}^{-s\tau_s}$} ;
\node[tf,label={above:Model}] (Gm)  [below =15pt of G]{$G_m(s)\mathrm{e}^{-s\tau_m}$};
\node[sum] (sum1) [left=10pt of Q]{};
\draw(sum1.north west) --(sum1.south east);
\draw(sum1.north east) --(sum1.south west);
\node[left = -3.5pt] at (sum1.center){};
\node[below = -2.75pt] at (sum1.center){-};


\node[sum] (sum2) [right=15pt of G]{};
\draw(sum2.north west) --(sum2.south east);
\draw(sum2.north east) --(sum2.south west);
\node[left = -3.5pt] at (sum2.center){};
\node[above = -2.5pt] at (sum2.center){};


\node[sum] (sum3) [right=37.5pt of Gm]{};
\draw(sum3.north west) --(sum3.south east);
\draw(sum3.north east) --(sum3.south west);
\node[left = -2.5pt] at (sum3.center){-};
\node[above=-2.5pt] at (sum3.center){};

\node[below=15pt of sum3](corner){};
\node[above=10pt of sum2](dd){};

\draw[-triangle 45](sum1.east) -- (Q.west);
\draw[-triangle 45](Q.east) -- (G.west) node[midway,above] {$u$} node[midway] (u) {};
\draw[-triangle 45](u.center) |-(Gm.west);
\draw[-triangle 45](G.east) -- (sum2.west);
\draw[-triangle 45](sum2.east) --++ (35pt,0pt) node[midway](output){} node[midway,above]{$y$};
\draw[-triangle 45](output.center) -- (sum3.north);
\draw[-triangle 45](Gm.east) -- (sum3.west) node[midway,above]{$y_m$};
\draw (sum3.south) -- (corner.center);
\draw[-triangle 45] (corner.center) -| (sum1.south);
\draw[-triangle 45] (dd.center) -- (sum2.north) node[midway,right]{$d$};
\draw[triangle 45-] (sum1.west) --++(-15pt,0pt) node[midway,above]{$r$};
\end{tikzpicture}}
    \vspace{-20pt} 
    \caption{The Internal Model Control scheme, with reference $r$, system input $u$ generater by the controller, measured system output $y$, and modelled system output $y_m$.}
    \label{fig:IMC_scheme}
\end{figure}
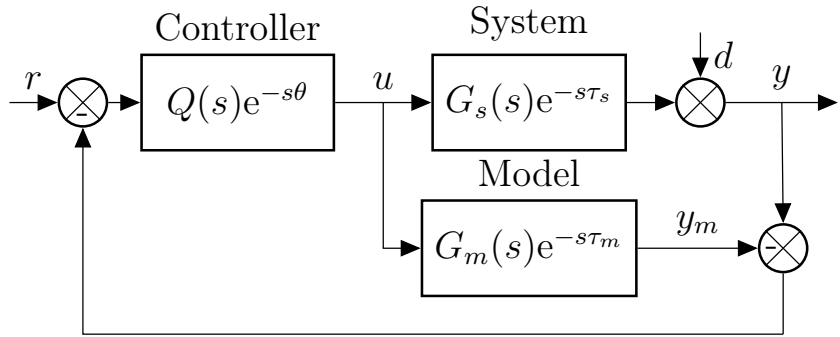

As shown in \figref{fig:IMC_scheme}, in the IMC control feedback arrangement, the system model is incorporated in the closed-loop. The controlled system is expressed by the transfer function $G_s(s)$ paired with time delay $\tau_s$. Similarly, the model of this system is given by the pair $G_m(s)$ and delay $\tau_m$, but they are not necessarily equal to that of the system. Finally, the controller consists of the transfer function $Q(s)$ and the time delay $\theta$ to be tuned during the control synthesis. 
Note that, IMC framework can be viewed as a particular case of the famous \emph{Youla-Ku\v{c}era parametrization} \cite{kuvcera2011method} (also known as Q-parameterization) obtained when the considered systems are stable systems. Hence, in a control scenario where the system assumptions made in this paper fail, one can refer to the general framework expressed by this parameterization.  However, in the particular case where our assumptions hold, a stabilizing controller parameterized by a stable and proper transfer function $Q(s)$ can be rearranged to yield the IMC scheme where the Q parameter becomes the IMC controller. As a result, stability of the closed-loop is implied by a stable and proper IMC controller which motivates the adoption of IMC framework for the systems we consider. Nevertheless, the parameterization does not shed light on how this parameter should be selected. From this perspective, the subsequent work presented below can be regarded as a way of choosing this parameter.

The \emph{sensitivity transfer function} $S(s)$ of this scheme is given by

\begin{equation}
    S(s) =\frac{Y(s)}{D(s)}= \frac{1 - Q(s)G_m(s) \euler^{-s (\tau_m + \theta)}}{1 + Q(s)(G_s(s) \euler^{-s (\tau_s + \theta)} - G_m(s) \euler^{-s (\tau_m + \theta)})}.
    \label{eq:true_sensitivity}
\end{equation}

Without any assumptions, it corresponds to a retarded time delay system with infinitely many zeros and poles since the numerator and the denominator  can be reduced to quasi-polynomials after some few calculations. Nevertheless, under an ideal case given by the following definition, it nominally becomes finite-dimensional:

\begin{definition}[Ideal Configuration]

In the ideal configuration i.e. when $G_s(s)=G_m(s)=G(s)$ and $\tau_s = \tau_m = \tau$, the infinite-dimensional sensitivity \eqref{eq:true_sensitivity} of the {IMC} scheme shown in \figref{fig:IMC_scheme} reduces to the finite-dimensional ideal sensitivity 
\begin{equation}
     S(s) = 1 -Q(s) G(s) \euler^{-s(\tau+\theta)}.
     \label{eq:ideal_sens}
\end{equation}
\end{definition}

\begin{remark}
    An important observation is that, whether the ideal configuration holds or not, the numerator in \eqref{eq:ideal_sens} remains unchanged and depends on the identified model and the controller. Hence, the regulation properties of the closed-loop are invariant to system/model mismatch, which motivates to use the ideal configuration to find the conditions the controller needs to satisfy.
\end{remark}

\subsection{Filter design through reference sensitivity}
As pointed out by IMP, in order to achieve the control goal, the controller should essentially result in placing the poles of the signal model \eqref{eq:finite_signal_model} as the zeros of the closed-loop sensitivity transfer function $S(s) = \frac{Y(s)}{D(s)}$, i.e.  
\begin{equation}
    S(0)= S(\mathrm{j} \omega_i) = 0.
    \label{eq:sensitivity_cond}
\end{equation} 
\begin{proposition}
Under the \emph{ideal configuration}, i.e., when the model and the system perfectly match, the controller structure constructed as
\begin{equation}
    Q(s) = \frac{1}{G(s)} F(s) ,
    \label{eq:controller}
\end{equation}
where $F(s)$ is a low-pass filter with relative degree $n_r$, satisfying $n_r\ge\beta-\alpha$, and with properties 
\begin{eqnarray}
    F(0) =\left| F(\mathrm{j} \omega_i) \right| &=& 1,  \label{eq:F0} \\
    \angle F(\mathrm{j} \omega_i)   - \omega_i(\tau+ \theta) &=& 2\pi h, \ h \in \mathbb{Z}, \label{eq:Fjw}
\end{eqnarray}
satisfies periodic control condition \eqref{eq:sensitivity_cond}.
\end{proposition}
\noindent\textbf{Proof.}\
Substituting controller structure \eqref{eq:controller} into ideal sensitivity \eqref{eq:ideal_sens} yields
\begin{equation}
    S(s) =1 - F(s) \euler^{-s(\tau+\theta)}.
    \label{eq:sens_filter}
\end{equation}
The required properties for the filter are obtained by reflecting condition \eqref{eq:sensitivity_cond} to \eqref{eq:sens_filter}. $\blacksquare$
\begin{remark}
    Since systems with non-minimum phase zeros are excluded from the study, the inverse of $G(s)$ can be directly incorporated into the controller, rendering point-wise approximation of the inverse of $G(s)$ unnecessary. Hence, the transient behavior of the closed-loop is set dominantly by the filter, simplifying the design steps. 
\end{remark}

 In what follows, the filter $F(s)$ in the controller structure is obtained by going backward from a chosen reference sensitivity prescribed as
\begin{equation}
    S_{\mathrm{ref}}(s) = \frac{z(s)}{p(s)} = \frac{s  \prod_{i=1}^k (s^2+\omega_i^2) \prod_{i=1}^m (s-z_i)}{\prod_{i=1}^n (s-p_i) },
    \label{eq:sens_poly}
\end{equation}
where $p_i\in\mathbb{C}^{-}$, $ \omega_i \in \mathbb{R}^{+}$ and $z_i \in\mathbb{C}$ denote the poles, the targeted frequencies and the additional zeros, respectively. The motivation behind choosing this as the reference sensitivity stems from the fact that $\eqref{eq:sens_poly}$ generally represents all regular stable transfer functions that have the required zeros for tracking/rejection on the imaginary axis at values corresponding to the targeted harmonic frequencies. 


\subsection{Filter structure}

Any filter-delay combination satisfying conditions \eqref{eq:F0}-\eqref{eq:Fjw} leads to a potential realization of the controller \eqref{eq:controller}, provided that the controller is causal. This fact leads to various different approaches to exist for the decision of these components. For instance, in our previous work \cite{yuksel2023harmonic}, an analytical filter design based on the combination of a second- and a first-order filter was proposed for each targeted frequency and was shown to be capable of suppressing several frequencies simultaneously when combined. Alternatively, in \cite{yuksel2023distributed}, a design method based on replacing the lumped delay $\theta$ in the controller with a distributed delay was proposed. The distributed delay had structure $\frac{1}{s} \sum_{i=0}^{N} a_i \euler^{-s\theta_i}$ and was shaped by tuning the gains $a_i$ rather than $\theta_i$. However, the previous proposed approaches relied on the assumption that the control system had an inner control loop that made the system approximable by a first-order model with time delay. With the following proposition, a controller can be designed directly using the plant model expressed as general as in \eqref{eq:G}.

\begin{proposition}
 Let $\omega_b$ be the so-called \emph{base frequency}, such that $\omega_b \leq \omega_i$ and $ \gamma:= \frac{\omega_i}{\omega_b} \in \mathbb{Z}, \  \forall i = 1, ..., k $. Based on the ideal configuration with sensitivity $S(s)$ given by \eqref{eq:ideal_sens}, the filter in the proposed controller \eqref{eq:controller} constructed as 
    \begin{equation}
        F(s)= \frac{p(s)-z(s)}{p(s)},
        \label{eq:ideal_filter}
    \end{equation} where $p(s)$ and $z(s)$ correspond to the denominator and numerator of the reference sensitivity $S_{ref}(s)$ in \eqref{eq:sens_poly}, respectively, satisfies condition \eqref{eq:sensitivity_cond} for the ideal sensitivity $S(s)$ provided that the controller delay is
    \begin{equation}
      \theta = \frac{2\pi l_b}{\omega_b}-\tau \ge 0,
      \label{eq:theta}
  \end{equation}
  where $l_b = \Big\lfloor\frac{\tau\omega_b}{2\pi}\Big\rfloor +1.$ 
\end{proposition}

\noindent\textbf{Proof.}\
Substituting the filter \eqref{eq:ideal_filter} to the ideal sensitivity \eqref{eq:sens_filter}, it reads as 
\begin{equation}
    S(s) = 1 - \frac{p(s)-z(s)}{p(s)}e^{-s(\theta+\tau)}.
    \label{eq:sens_with_pz}
\end{equation}
Since, by definition of $z(s)$ in \eqref{eq:sens_poly}, $F(0)=F(\mathrm{j} \omega_i)=1$, ideal sensitivity in \eqref{eq:sens_with_pz} at targeted frequencies becomes 
\begin{equation}
    S(\mathrm{j} \omega_i) = 1 - \euler^{-\mathrm{j} \omega_i(\theta+\tau)}.
\end{equation}
Requiring condition \eqref{eq:sensitivity_cond} to hold leads to \eqref{eq:theta}. Note that if $\omega_b(\tau+\theta)$ is an integer multiple of $2 \pi$ then so as $\omega_i(\tau+\theta)= \gamma \omega_b (\tau+\theta).$ $\blacksquare$

\begin{remark}
In terms of polynomials $p(s)$ and $z(s)$ of the reference sensitivity \eqref{eq:sens_poly}, the sensitivity of the {IMC} scheme \eqref{eq:sens_filter} can be expressed as
\begin{equation}
    S(s) =\frac{p(s)\left(1-\euler^{-s(\tau+\theta)}\right)+z(s)\euler^{-s(\tau+\theta)}}{p(s)}.
    \label{eq:sens_filter3}
\end{equation}
Thus, next to preserving the imaginary axis zeros, it preserves the poles $p(s)$. Due to the quasi-polynomial nature of the numerator, compared to \eqref{eq:sens_poly}, it has infinitely many additional zeros. The references match each other for $\tau+\theta=0$.
\end{remark}

\subsection{Quasi-optimal filter design in state-space form}

Proposition 2 suggests a general filter structure for which the required zeros for periodic control are placed to the sensitivity simply through $z(s)$ such that $z(0)=z(\mathrm{j} \omega_i)=0$ and the stability of this sensitivity is ensured by setting $p(s)$ as Hurwitz. Nevertheless, for the controller to be physically implementable, the relative order of the filter has to be no less than the relative degree of the system model, i.e., $n_r \geq \beta-\alpha$. This requirement on the relative degree can be achieved using the auxilary zeros denoted by $z_m$ and can be easily satisfied when formulated in state-space. Note that forming the filter and therefore the controller in the state-space representation is beneficial also from the implementation point of view. The multi-harmonics compensation regularly leads to high order of the filter \eqref{eq:ideal_filter}, the implementation of which directly in the transfer function form can be numerically risky due to forming the higher order polynomials from poles and zeros.

\begin{proposition}
 Let $\Sigma(A,B,C,D)$ be a minimal state-space realization of the reference sensitivity in \eqref{eq:sens_poly} such that $\det(s I -A )=p(s)$ and $C(\mathrm{j} \omega_i I-A)^{-1}B = - D$. Then, for a fixed output matrix $C$ and $D=1$, the filter given by \eqref{eq:ideal_filter} that targets harmonic frequencies $\omega_i,i=1...k$ with desired relative degree $n_r$ has a unique input matrix $B$ satisfying
 \begin{equation}
    \begin{bmatrix}
        CA^{-1} \\
        \Re(C(\mathrm{j} \omega_1 I -A)^{-1}) \\
        \Im(C(\mathrm{j} \omega_1 I -A)^{-1}) \\
        \vdots \\
        \Re(C(\mathrm{j} \omega_k I -A)^{-1}) \\
        \Im(C(\mathrm{j} \omega_k I -A)^{-1})
    \end{bmatrix}_{(2k+1) \times n}  B = 
    \begin{bmatrix}
        -1 \\
        -1 \\ 
        \ 0 \\
        \vdots \\
        -1 \\
        \ 0
    \end{bmatrix}
    \label{eq:zero_condition}
\end{equation}
and \begin{equation}
    \begin{bmatrix}
        C \\
        CA \\
        \vdots \\
        CA^{(n_r-2)}  
    \end{bmatrix}_{m \times n}  B = 
\begin{bmatrix}
    0 \\
    0 \\
    \vdots \\
    0
\end{bmatrix}.    
\label{eq:nr_condition}
\end{equation} 
Consequently, the state-space representation of the filter $F(s)$ given by \eqref{eq:ideal_filter} reads as $\Sigma(A,B,-C,0)$.

\end{proposition}

\noindent\textbf{Proof.}\
It is easy to observe that \eqref{eq:ideal_filter} can be expressed as $F(s)=1- S_{\mathrm{ref}}(s)$ which gives the 
state-space form of the filter as $\Sigma(A,B,-C,1-D)$. 
Thus, in order to have $n_r \geq 1$, it is necessary 
that $D=1$. Subsequently, this implies that \eqref{eq:sens_poly} has to correspond to a biproper transfer function, i.e. $n=2k+1+m$. Based on this 
observation, for a fixed $C$, one can derive $n$ linear equations to uniquely find $B$. The first $2k+1$ equations expressed by \eqref{eq:zero_condition} are obtained by imposing \eqref{eq:sensitivity_cond}. The rest $n-2k-1$ equations are introduced by the condition \eqref{eq:nr_condition} for the relative degree, which follows from letting the occurring derivative terms of the state and the input have zero coefficients after taking the derivative of the output $n_r$ times. Notice that, from \eqref{eq:nr_condition}, the number of auxiliary zeros is related to the desired relative degree by $m=n_r-1$.  Hence, based on the specified $k$ and $n_r$, we can form $n$ equations from which $B$ can be found uniquely. $\blacksquare$
\\

Proposition 3 clearly 
shows the 
roles of the zeros: some zeros are placed to achieve periodic control, and the rest are placed to make the controller physically realizable. 

Additionally, it is necessary to place the poles of the filter, i.e. the eigenvalues of the matrix $A$. 
%
For achieving the periodic control task, any position of the poles on the open left-half plane is possible. However, their unsuitable distribution may have negative consequences on transients of the control system. Therefore, we adopt the standard LQR approach \cite{lewis2012optimal} to place most of the poles. 

%
Rearranging the filter structure as
\begin{equation}
    F(s) = \frac{p(s)-z(s)}{p(s)} = \frac{\frac{p(s)-z(s)}{z(s)}}{1+ \frac{p(s)-z(s)}{z(s)}}
\end{equation}
and introducing the new variable $\eta(s):=p(s)-z(s)$, the filter structure can be considered as a closed-loop system with a controller $\eta(s)$ and a marginally stable plant $\frac{1}{z(s)}$. First, in the following proposition, we derive $\eta(s)$ for a signal model \eqref{eq:finite_signal_model} only, i.e. considering $\frac{1}{z(s)}=V(s)$ and thus $z(s)$ being free of auxiliary zeros. The proposed approach is directly applicable for designing LQR optimal filter with relative degree $n_r=1$. Subsequently, we address the generalization for more common filter structure with $n_r\ge1$, leading to quasi-optimal solution.



\begin{figure}[ht]
    \centering
    \scalebox{1.2}{\begin{tikzpicture}[
tf/.style = {rectangle, draw=black, fill=white,  thick, minimum width=60pt, minimum height=30pt},
tf_small/.style = {rectangle, draw=black, fill=white,  thick, minimum width=30pt, minimum height=20pt},
sum/.style = {circle,draw=black,fill=white, thick, minimum size=10pt}]


\node[tf,label={above:Signal Model}] (R){$\dot x_R = A_R x_R + B_R u_R$};
\node[tf_small] (Feedback_Eta) [below = 5 pt of R] {$K$};
\node[] (Corner_Node) [right=10 pt of R] {};
\node[] (Corner_Below) [below=40pt of Corner_Node] {};
\node[] (Output) [right =25 pt of Corner_Below ] {$y_f$};
\node[] (Output_channel) [left = 15 pt of Feedback_Eta] {};

\node[sum] (sum_filter) [left = 10 pt of R] {} ;
\draw(sum_filter.north west) --(sum_filter.south east);
\draw(sum_filter.north east) --(sum_filter.south west);
\node[below = -3.5 pt] at (sum_filter.center){-};

\draw[-triangle 45](sum_filter.east) -- (R.west);
\draw[-](R.east) -- (Corner_Node.center) node[above] (x_r) {$x_R$};
\draw[-triangle 45](Corner_Node.center) |- (Feedback_Eta.east);
\draw[-triangle 45](Feedback_Eta.west) -| (sum_filter.south);
\draw[-triangle 45] (Output_channel.center) |- (Output);
\draw[triangle 45-] (sum_filter.west) -- ++ (-20pt,0pt) node[midway] (u) {} node[left, above] (u_f) {$u_f$}  ;

\draw[thick,dotted] ($(u)+(0pt, -60pt)$) rectangle ($(Output)+(-25pt, +90pt)$) node[below= 10pt,left] {$F(s)$};

\end{tikzpicture}}
    \vspace{-20pt} 
    \caption{The filter structure in state feedback form for $n_r=1$.}
    \label{fig:filter_state}
\end{figure}
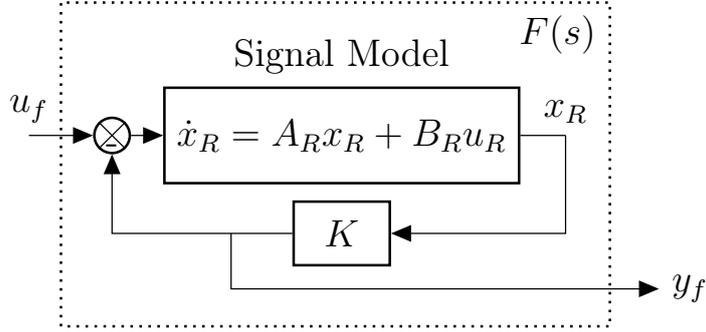

\begin{proposition}
    Let $\Sigma[A_R, B_R,C_R,D_R]$ be a minimal state-space realization of the signal model $V(s)$ given by \eqref{eq:finite_signal_model}.
    Let its state, input and output vectors be denoted by $x_R$, $u_R$ and $y_R$, respectively. Additionally, let $K_\eta$ be the vector containing the coefficients of $\eta(s)$ such that $\eta(s) = K_{\eta}\left[1 \quad s \quad ... \quad s^{n-2} \quad s^{n-1} \right]^\intercal$. The optimal coefficients for $\eta(s)$ that minimize the quadratic cost function
    \begin{equation}
        \int_{0}^{\infty} x_R^\intercal \mathcal{Q} x_R + u_R^\intercal \mathcal{R} u_R \,\mathrm{d}t
        \label{eq:cost}
    \end{equation}
    is given by
    \begin{equation}
        K_\eta =  K O_R^{-1} ,
        \label{eq:gain_cond}
    \end{equation}
    where $O_R$ is the observability matrix of the signal model and $K$ is the optimal state-feedback gain minimizing the cost function for the filter in the state-feedback configuration shown in \figref{fig:filter_state}. Additionally, the resulting optimal filter with the optimal $\eta(s)$ has a relative degree $n_r=1$.
    
\end{proposition}

\noindent\textbf{Proof.}\
Based on the state-space representation of $\frac{1}{z(s)}=V(s)$ we can 
express the filter as the interconnection of
\begin{eqnarray}
    sX_R(s) &= A_R X_R(s) + B_R U_R(s),\\
    Y_R(s) &= C_R X_R(s) + D_R U_R(s),
\end{eqnarray}
with $U_R(s) = U_f(s) - \eta(s) Y_R(s)$ and $Y_f(s) = \eta(s) Y_R(s)$.
Since we seek a $\eta(s)$ that stabilizes the system, without loss of generality, we can set $u_f(t)=0$. This yields the feedback effect to be expressed as
\begin{eqnarray}
     U_R(s) &=& \eta(s) Y_R(s) \\
     &=& K_{\eta} \left[1 \quad s \quad ... \quad s^{n-2} \quad s^{n-1} \right]^\intercal Y_R(s),
\end{eqnarray}
which in time-domain corresponds to  
\begin{equation}
    u_R(t) = K_{\eta}   \begin{bmatrix}
        y_R(t) \\
        \dot y_R(t) \\
        \vdots \\
        y_R^{(n-1)}(t)
    \end{bmatrix} = K_{\eta} \underbrace {\begin{bmatrix}
        C_R \\
        C_R A_R \\
        \vdots \\
        C_R A_R^{n-1}
    \end{bmatrix}}_{O_R} x_R(t).
    \label{eq:state_feedback}
\end{equation}
Note that, due to having the relative order of $\frac{1}{z(s)}$ equal to $n-1$, we have $C_R A_R^r B_R = 0$ for $r = 0,1 \hdots n-2$, which in return gives rise to the observability matrix $O_R$. Notice that the final form of \eqref{eq:state_feedback} corresponds to the state feedback with gains $K := K_{\eta} O_R$ and allows the closed-loop to be represented as in \figref{fig:filter_state}. Hence, using the standard {LQR} approach, one can find $K$ that optimally stabilizes the closed-loop with respect to cost \eqref{eq:cost}. Since the state-space realization of $\frac{1}{z(s)}$ is minimal, $O_R^{-1}$ exists and can be used to find $K_\eta$ as in \eqref{eq:gain_cond}. Lastly, since the degree of $\eta(s)$ is related to the relative degree of the filter through $\mathrm{deg}(\eta(s)) = n - n_r$, the resulting filter has $n_r=1$. $\blacksquare$
\\

In accordance with Proposition 4, the LQR optimal filter \eqref{eq:ideal_filter} with $n_r=1$ is given by
\begin{eqnarray}
    \dot x_R(t)  &=& \left(A_R - B_R K \right) x_R(t) + B_R u_f(t) ,\label{eq:filter_state1}\\
    y_f(t) &=& K x_R(t) .
    \label{eq:filter_state2}
\end{eqnarray}
For filters with $n_r>1$ resulting to $z(s)$ with auxiliary zeros, the approach proposed in Proposition 4 cannot be applied directly. Since the high-order terms of $\eta(s)$ are set to zero to achieve the desired relative degree, the $K$-feedback is not taken from the full state.
A practical way to solve the task is to expand the dynamic matrix of the filter to 
%
\begin{equation}
    A = \begin{bmatrix}
        A_R -B_RK & 0 \\ 
        0  & A_{\mathrm{rel}}
    \end{bmatrix}, \\ 
    \label{eq:matrix}
\end{equation}
where $A_{\mathrm{rel}} \in \mathbb{R}^{(n_r-1) \times (n_r-1)}$ is a Hurwitz matrix with $n_r-1$ pre-selected poles, possibly located sufficiently far from the spectrum obtained by LQR design so that their effect on the overall dynamics is minimal and the filter performance is close to optimal. 

Before moving forward to forming the overall {IMC} controller, let us outline the overall procedure of designing the state-space realization $\Sigma(A,B,-C,0)$ of the filter $F(s)$:
\begin{enumerate}
    \item According to Proposition 4, select $\mathcal{Q}$ and $\mathcal{R}$ and obtain $K$ by solving LQR problem \eqref{eq:cost} for the signal model \eqref{eq:finite_signal_model} with $A_R, B_R$.
    \item Pre-select the $n_r-1$ poles of the matrix $A_{\mathrm{rel}}$ and form the filter dynamics matrix $A$ as in \eqref{eq:matrix}.
    \item By Proposition 3, for a fixed $C$ determine $B$ by solving \eqref{eq:zero_condition}-\eqref{eq:nr_condition}. Note that, for convenience, the matrix $C$ can be chosen to consist of only ones i.e. $C= \left[1 \ 1 \ ... \ 1  \right]$ to ensure that all states reflect on the system output regardless of the structure of matrix $A$. 
\end{enumerate}

\subsection{State-space representation of IMC controller}

The overall {IMC} controller $Q(s)\euler^{-s\theta}$ is given in the state-space form
\begin{eqnarray}
    \dot x_Q(t) &=& A_Q x_Q(t) + B_Q u_Q(t), \label{IMCfinal1} \\
    u(t) &=& C_Q x_Q(t-\theta) + D_Q u_Q(t-\theta), 
    \label{IMCfinal2}
\end{eqnarray}
where $u_Q(t)$ and $u(t)$ are the input and the output of the controller respectively. 
From the possible state-space representations, a straightforward formulation of the system matrices can be obtained as follows. Let $Q(s)=F(s)G_\alpha(s)G_\beta(s)$, where $ G_{\alpha}(s) = ( a_{\alpha}s^\alpha+a_{\alpha-1}s^{\alpha-1}+ ... + a_1s+a_0)^{-1}$ and $G_\beta(s)=b_{\beta}s^\beta+b_{\beta-1}s^{\beta-1}+ ... + b_1s+b_0$, then
\begin{eqnarray}
A_Q &=& \Tilde{A} = \begin{bmatrix}
        A & 0 \\ 
        -B_\alpha C & A_\alpha
    \end{bmatrix}, \\ 
B_Q &=& \Tilde{B} =\begin{bmatrix}
        B \\
        0
    \end{bmatrix}, \\
       C_Q &=& \begin{bmatrix}
        b_{\beta} & b_{\beta-1} & \dots & b_1 & b_0
    \end{bmatrix} 
    \begin{bmatrix}
        \Tilde{C}\Tilde{A}^\beta \\
        \Tilde{C}\Tilde{A}^{\beta-1} \\
        \vdots \\ 
        \Tilde{C}
    \end{bmatrix},\\
    D_Q &=& b_{\beta} \Tilde{C} \Tilde{A}^{(\beta-1)} \Tilde{B},
\end{eqnarray}
where $\Sigma(A_\alpha,B_\alpha,C_\alpha,D_\alpha)$ is a minimal state-space realization of
$G_{\alpha}(s)$ and $\Sigma(\Tilde{A}, \Tilde{B} ,\Tilde{C} ,0)$ is a  representation of $F(s)G_\alpha(s)$, with $\Tilde{C} =[-D_\alpha C \quad C_\alpha]$. 
Subsequently, combining $\Sigma(\Tilde{A}, \Tilde{B} ,\Tilde{C} ,0)$ with $G_\beta(s)$ yields the $C_Q$ and $D_Q$ controller matrices. 

\subsection{Robustness against model/plant mismatch}

The ideal configuration does not apply to physical realizations, since in practice there will always be an unaccounted part of the system dynamics in the model. However, due to the internal model in the IMC scheme and the fact that the spectral abscissa of the time delay system continuously varies with respect to the coefficients of the characteristic equation \cite{michiels2014stability}, the closed-loop inherently attains a certain level of robustness against mismatches between the system and its model.
Nevertheless, the well-known \emph{Small Gain Theorem} \cite{skogestad2005multivariable} can be used to investigate and improve the robust closed-loop performance with the proposed filter in a more quantitative manner: 

\begin{proposition}
Consider the mismatch between the system and its model $\Delta(s):=G_s(s)\euler^{-s\tau_s} - G_m(s) \euler^{-s\tau_m}$. Based on the Small-Gain Theorem, a sufficient condition to ensure the stability of the closed-loop scheme with sensitivity \eqref{eq:true_sensitivity}, is given by 
    \begin{equation}
        \left| \left| \Delta(j\omega) Q(j\omega)  \right| \right|_\infty < 1,
        \label{eq:small_gain}
    \end{equation} 
    where $Q(s)$ is the IMC controller as in \eqref{eq:controller}.
    \label{mismatch}
\end{proposition}

\noindent\textbf{Proof.}\ The close-loop mapping from $R(s)$ to $Y(s)$ can be found as 

\begin{equation}
    Y(s) = G_s(s) Q(s) \euler^{-s(\tau_s+\theta)} \frac{1}{1 + \Delta(s) Q(s) \euler^{-s\theta}} R(s). 
\end{equation}
Notice that, $G_s(s) Q(s) \euler^{-s(\tau_s+\theta)}$ is stable by construction. Hence, the closed-loop transfer function is stable if and only $\frac{1}{1 + \Delta(s) Q(s) \euler^{-s\theta}}$ is stable. Regarding the latter expression as a closed-loop transfer function on its own and then applying the Small Gain Theorem gives \eqref{eq:small_gain}. $\blacksquare$
\\

Nevertheless, note that the term $\Delta(s)Q(s)$ in Proposition \ref{mismatch} generally corresponds to an \emph{infinite-dimensional system}. Therefore, its $\mathcal{H}_\infty$-norm should be evaluated with this fact in mind. Since the systems and controllers considered are SISO, one can refer to the method proposed in \cite{gumussoy2015computation} for an effective way to calculate the $\mathcal{H}_\infty$-norm. In the case where the perturbation that causes the mismatch is known, one can also use the spectral distribution to assess the stability in a root-locus fashion. For more insight into the effects of plant/model mismatch on control performance, see \cite{badwe2010quantifying}.

\section{Experimental case study validation}
\begin{figure}[ht]
    \centering
    \includegraphics[width=0.8\columnwidth]{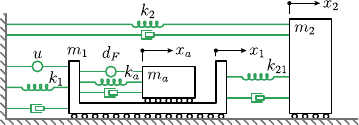}
    \caption{Scheme of the laboratory setup}
    \label{fig:setup}
\end{figure}

The proposed controller design and its performance are experimentally tested on a sixth-order system corresponding to the serially connected mass-spring-damper with two actuators depicted in \figref{fig:setup}. The goal is to control the position $x_2$ of the mass $m_2$ by the actuator $u$ with output feedback despite the oscillations caused by the actuator $d_F$ and the artificially added input delay of $\tau=\SI{0.2}{s}$ located at the input $u$. The disturbance enters the configuration in the form of force $d_F$ which is a sawtooth with period $T=\SI{0.5}{s}$, i.e. with base frequency $\omega_b=\SI[parse-numbers=false]{4\pi}{rad/s}$ ($f_b=\SI{2}{Hz}$). 

 \begin{figure}[ht]
     \centering
     \includegraphics[width=1\columnwidth]{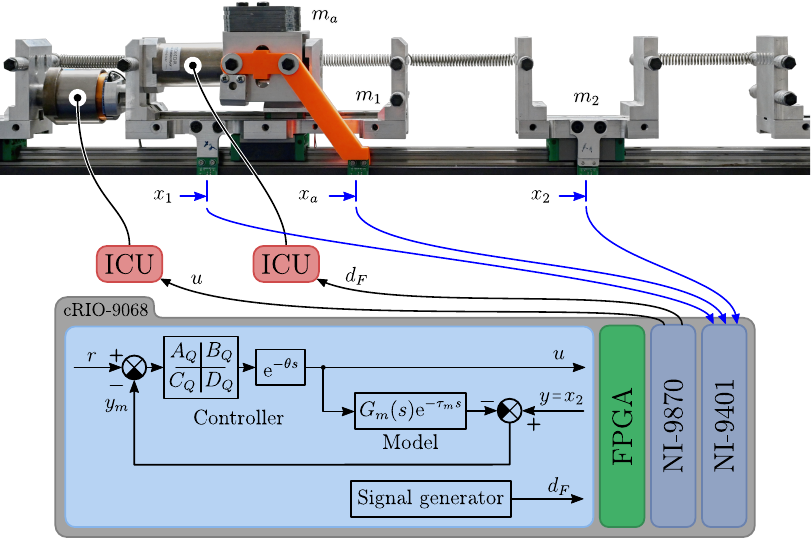}
     \caption{Mechatronic implementation of the laboratory setup}
    \label{fig:control_loop}
\end{figure}

\subsection{Instrumentation and mechatronic design}

The mechatronic implementation of the laboratory setup sketched in \figref{fig:setup} is shown in \figref{fig:control_loop}. The setup is made up of three carts denoted by their masses $m_1$, $m_2$ and $m_a$. The carts $m_1$ and $m_2$ are mounted on the base rail and the cart $m_a$ is placed on a rail mounted at the top of $m_1$. The carts are free to move along their axis given by their respective rails thanks to industrial ball bearings. The interconnection of the masses is achieved through the springs connected, as illustrated in the scheme in \figref{fig:setup}. For damping, no exclusive components are used; the illustrated damping elements capture those yielded by inherent damping of the springs in combination with viscous friction. Note that the non-linearity brought about by dry friction is neglected and serves as model uncertainty.  

The displacements of the carts are measured via a multipole magnetic strip, located just below the ground rail, with a resolution of $\SI{25}{\micro\metre}$. Each cart is equipped with an AMS AS5304 incremental position sensor with Hall elements reading a quadrature signal. Actuation of the setup is achieved through two linear voice-coil motors (LVCM): The Moticont LVCM-032-076-20 placed between $m_1$ and $m_a$ generates the periodic disturbance $d_F$. The LVCM AVM40-20-0.5 by Akribys installed between ground and $m_1$ generates the control actuation of the system corresponding to the manipulated variable $u$.  Both LVCMs are driven by two custom-made Industrial Control Units ({ICU}s) from PearControl.

The discrete-time form of the {IMC} controller \eqref{IMCfinal1}-\eqref{IMCfinal2} is obtained by the zero-order hold method with a sampling of $\SI{1}{\kilo\hertz}$ and is implemented by {LabVIEW\textsuperscript{\texttrademark}} to be performed by the CompactRIO controller (cRIO-9068). CompactRIO controller consists of a microprocessor and an FPGA module. The microprocessor computes the nominal forces $u$ and $d_F$. The FPGA module is used to (i) read the RS-422 quadrature incremental signals from position sensors using the NI-9401 digital input-output card, (ii) decode to increment or decrement the relative positions $x_2$, and (iii) command the ICUs to control the LVCMs. Both voice coil motors operate in the force regime. The nominal values of the actuation force $u$ and the disturbance force $d_F$ to be applied to the moving carts are transmitted from the serial card NI-9870 via RS-232 to the ICU. Both the reading of the quadrature signals together with its decoding and the transmission of the reference forces through RS-232 are also implemented in {LabVIEW\textsuperscript{\texttrademark}}.

\subsection{Parameter identification and output disturbance analysis}

\begin{figure}[t]
    \centerline{\includegraphics[width=0.9\columnwidth]{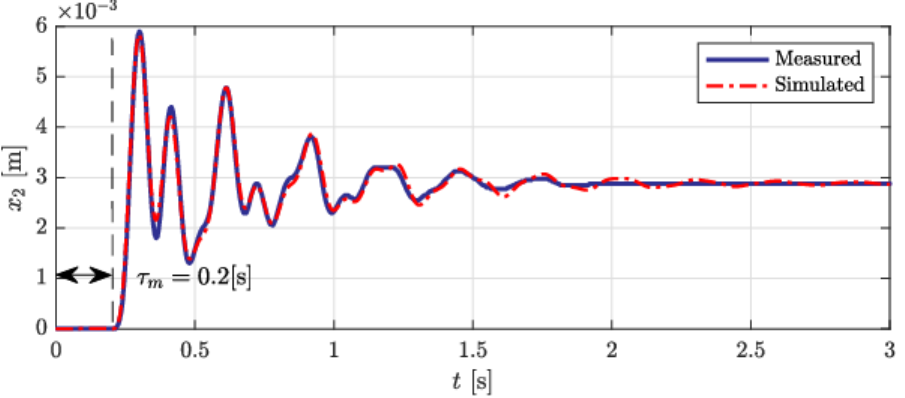}}
    \caption{Response of mass $m_2$ to a unit step input from $u$ initiated at $t=\SI{0}{\second}$. Notice that the effect of the added input-delay $\tau=\SI{0.2}{\second}$ is visible in the beginning.  }
    \label{fig:system_step}
\end{figure}

The known physical parameters of the  
experimental setup with the scheme in Fig. \ref{fig:setup} are the massess $m_1 = \SI{1.175}{\kilogram}$, $m_a = \SI{0.520}{\kilogram}$ and $m_2 = \SI{0.500}{\kilogram}$, and the stiffness coefficients of the springs $k_1 = \SI{1150}{Nm^{-1}}, k_{21}= \SI{749}{Nm^{-1}}$, $k_a = \SI{407}{Nm^{-1}} $, and $k_2 = \SI{711}{Nm^{-1}}$. However, the difficulty of identifying the damping induced by friction makes physics-based identification of the system challenging. For this reason, a model is derived based on the measured input-output signals of the original system. This approach is not only effective but also more suitable for industrial applications. Measurements for the identification procedure for model $G_\tau(s)$ are performed by recording the response of the position $x_2$$[\rm{m}]$ for a square wave given through input $u$$[\rm{N}]$ with an amplitude of $\SI{8}{N}$ and a period of $\SI{8}{s}$. Identification of the model $G_\tau(s)$ that maps the control actuation $u$ to the regulated position $x_2$ is achieved using MATLAB's \texttt{tfest} function, and is obtained in the form \eqref{eq:Gtau} 
comprising $G(s)$ given by \eqref{eq:G} with $\alpha=2$, $\beta=6$ and  parametric values  $b_0 = 3 \times 10^9\ \rm{s^{-6}}$, $b_1= 3.3 \times 10^7\ \rm{s^{-5}}$, $ b_2= 8.4 \times 10^6\ \rm{s^{-4}}$, $b_3 = 5.2 \times 10^4\ \rm{s^{-3}}$, $b_4 =5764\ \rm{s^{-2}}$, $b_5= 4.2\ \rm{s^{-1}}$, $b_6=1$, $a_0 = 1.031 \times 10^6\rm{mN^{-1}s^{-6}}$, $a_1= 4991\rm{mN^{-1}s^{-5}}$, $a_2=1258\ \rm{mN^{-1}s^{-4}}$ and the time delay $\tau = \SI{0.2}{s}$. The identified model is applied within the IMC scheme in \figref{fig:IMC_scheme}, considering $G_m(s)=G(s)$ and $\tau_m=\tau$. Note that the chosen $\beta$ captures the physical order of mass-spring-damper system in Fig. \ref{fig:setup} and the evaluated magnitudes for the coefficients are consistent with the above-stated physical values and units for the parameters of the setup. 

The effectiveness of the identification is demonstrated in \figref{fig:system_step} by comparing the measured response with the simulated response of the identified model. Notice that in the dynamically distinct part of the response, the match is almost perfect. However, the full accommodation at the equilibrium position of the cart $m_2$ takes longer for the simulation model. This is due to the slight dry friction forces present in the setup, which silence it sooner compared to the linear model with asymptotic behavior.  

\begin{figure}[tb]
    \centerline{\includegraphics[width=0.9\columnwidth]{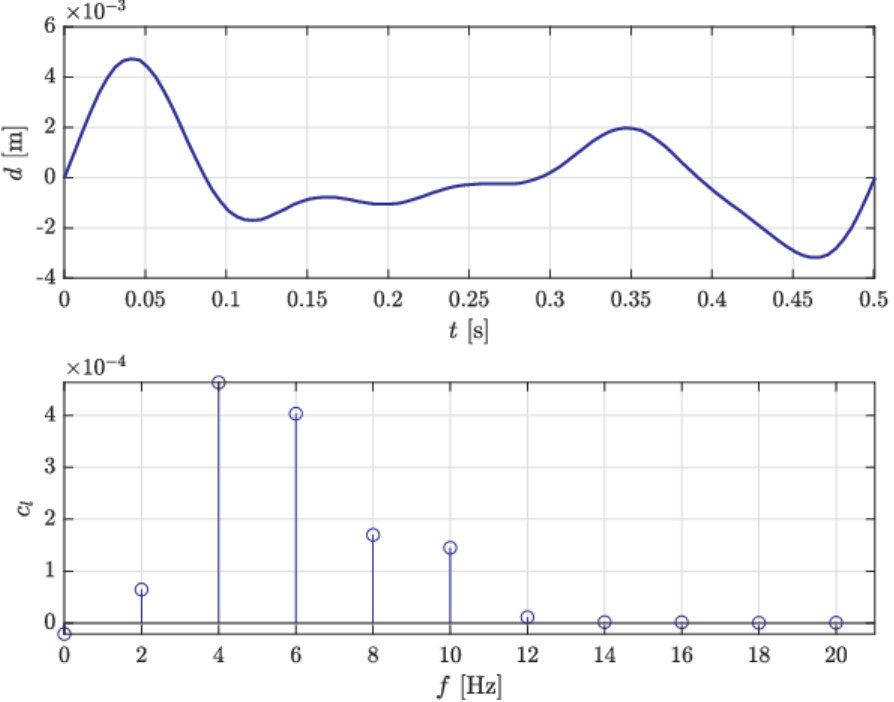}}
    \caption{({\bf Top}) A single period of the acting disturbance in time domain and ({\bf Bottom}) its composition in terms of harmonics as in \eqref{eq:Fourier}.}
    \label{fig:disturbance}
\end{figure}

Next, in \figref{fig:disturbance} we provide a projection of the $\SI{2}{Hz}$ sawtooth signal acting on the input $d_F$ to the output disturbance $d$. In addition to the period profile, the coefficients $c_l$ of the Fourier series expansion \eqref{eq:Fourier} are shown. As can be seen, the signal is composed of six dominant harmonics. In order to demonstrate the controller design capability and in order to eliminate even slight residual oscillations, we subsequently target $k=8$ harmonics $\omega_i=i\omega_b, i=1...k$.

\subsection{Controller design}
Since the identified model $G_\tau(s)$ has a relative order of 4, the relative order for the filter is chosen as $n_r=5$ to yield a causal controller with a relative order one. Thus, due to $k=8$, the order of the filter $F(s)$ is $n=21$ and the order of the {IMC} controller \eqref{IMCfinal1}-\eqref{IMCfinal2} is $n+\alpha=23$. 

First, the LQR problem \eqref{eq:cost} with selected $\mathcal{Q}=1000 \mathbf{I}_{17\times17}$ and $\mathcal{R}=1$ is solved using MATLAB's \texttt{lqr} function for the state-space representation $\Sigma[A_R, B_R,C_R,D_R]$ of the signal model \eqref{eq:finite_signal_model} built for the targeted frequencies, providing $K$. Next, the matrix $A_{21\times 21}$ given by \eqref{eq:matrix} is supplemented by $A_{\mathrm{rel}}$ matrix in Jordan canonical form with eigenvalues placed in proximity to $-100$, i.e., $\Re(\sigma(A_{\mathrm{rel}})) \approx -100 $. 
Prefixing $C_{1\times 21}=[1, 1, ... 1]$, subsequently $B_{21\times 1}$ is found using conditions \eqref{eq:zero_condition} and \eqref{eq:nr_condition}.
\begin{figure}[tbp]
    \centerline{\includegraphics[width=0.95\columnwidth]{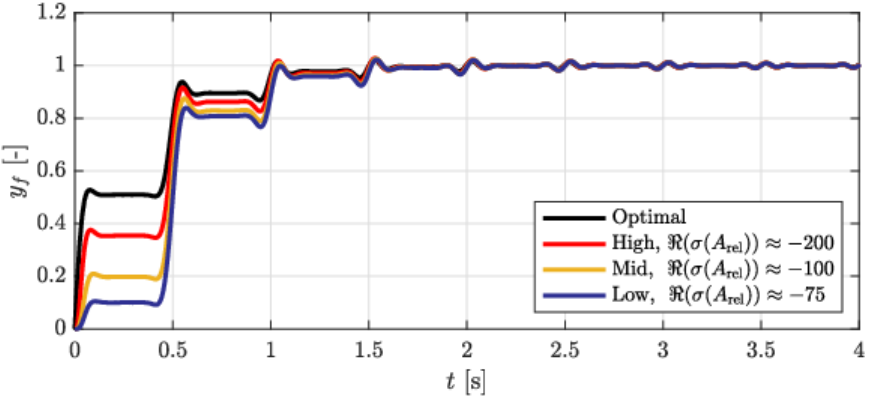}}
    \caption{Comparison of filters with $n_r=5$ but different $A_{\mathrm{rel}}$ with respect to optimal filter with $n_r=1$. "Low","Mid" and "High" imply the relative increase in $\left| \Re(\sigma(A_{\mathrm{rel}})) \right|$. The filter utilized in the experiment corresponds to the one with $\Re(\sigma(A_{\mathrm{rel}})) \approx -100 $.}
    \label{fig:comparison}
\end{figure}

The step response of the resulting filter $F(s)$ is shown in \figref{fig:comparison} in contrast to the optimal filter \eqref{eq:filter_state1}-\eqref{eq:filter_state2} and other expansions of the optimal filter with different $\Re(\sigma(A_{\mathrm{rel}}))$. As can be seen in the figure, placing the poles of $A_{\mathrm{rel}}$ further to the left in the complex plane makes the expanded filter behave similarly to that of the optimal one. Nevertheless, placing them too far can cause stiffness issues when discretizing the controller and prevent its physical realization. Therefore, a compromise between optimal behavior and physical realization must be made when placing the poles to account for the relative degree of the filter. For our setting, letting $\Re(\sigma(A_{\mathrm{rel}})) \approx -100 $ is a feasible solution that forms an acceptable trade-off. 

Consequently, the IMC controller \eqref{IMCfinal1}-\eqref{IMCfinal2} with the matrices formed as described in Section 2.4 and the additional controller delay $\theta=\SI{0.3}{s}$ obtained by \eqref{eq:theta}. 

\subsection{Spectral and frequency domain analysis}

\begin{figure}[t]
    \centering
    \includegraphics[width=1\columnwidth]{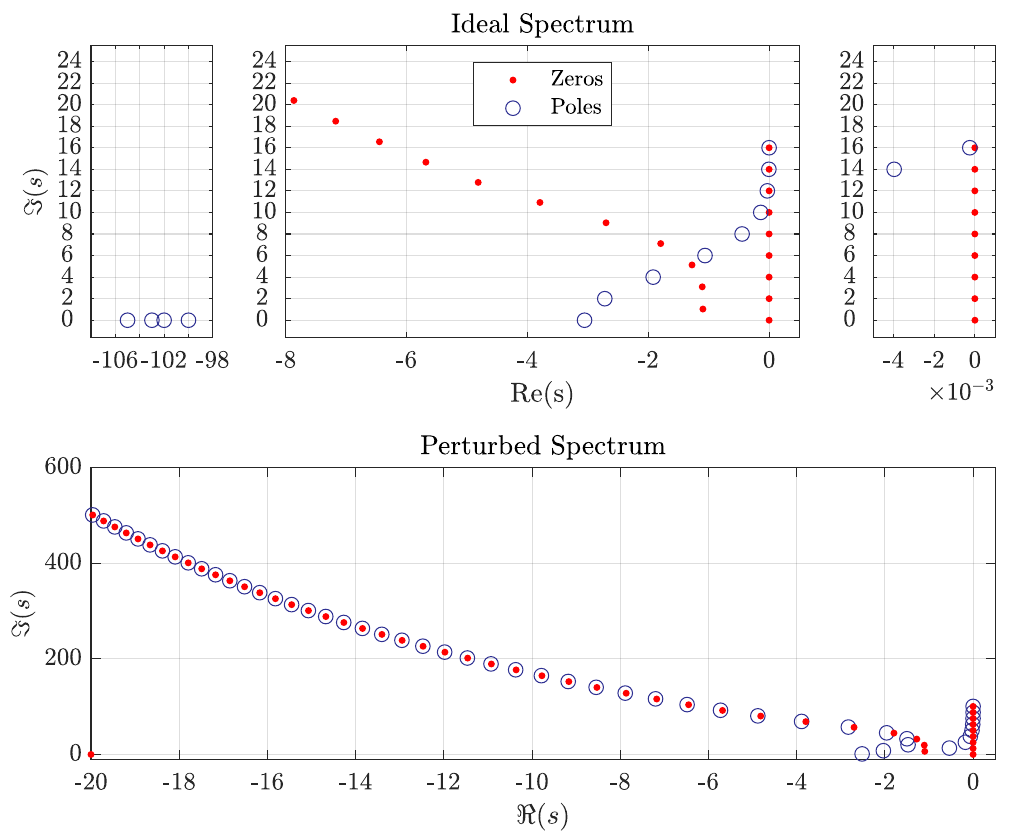}
    \caption{\textbf{(Top)} Pole-zero spectra of the ideal sensitivity \eqref{eq:ideal_sens} i.e. when $G_s(s)=G_m(s)$. \textbf{(Bottom)} Spectrum of the non-ideal sensitivity given by \eqref{eq:true_sensitivity} when system is perturbed as $G_s(s) = G_m(s) \frac{0.9}{0.05s+1}$. Notice the formation of a chain of infinite poles asymptotically converging to that of the zeros in contrast to the ideal spectrum.
    }
    \label{fig:spectrum}
\end{figure}

\begin{figure}[t]
    \centering
    \includegraphics[width=0.9\columnwidth]{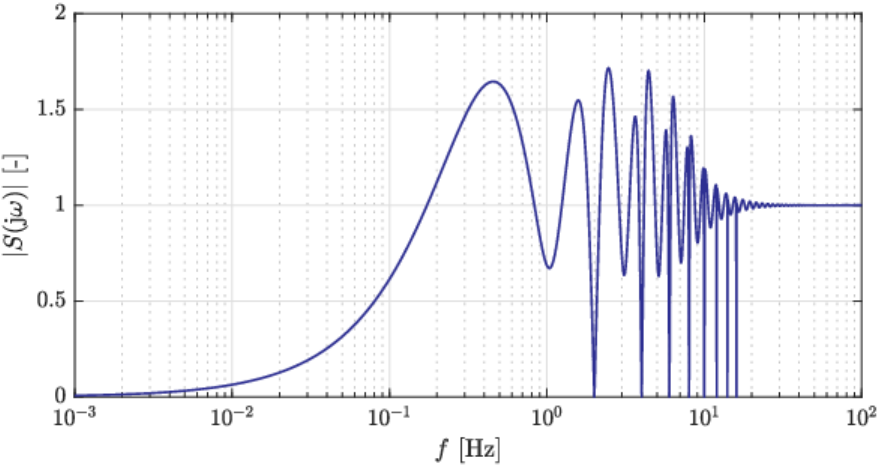}
    \caption{Frequency response of the sensitivity }
    \label{fig:bode}
\end{figure}

First, we analyze the spectral features of the resulting sensitivities, both ideal and perturbed, with the generated controller having the spectra of poles and zeros shown in \figref{fig:spectrum}, which were computed by the {QPmR} algorithm \cite{vyhlidal2009mapping}.
For the ideal sensitivity \eqref{eq:ideal_sens}, there are finitely many poles imposed within the design that are all in the stable region. The poles in the top middle figure correspond to those placed by the LQR, and the four poles seen in the top left are those induced by $A_{\mathrm{rel}}$, all together reaching to a total of 21 poles. On the other hand, there are infinitely many zeros that form a chain due to the fact that the numerator in \eqref{eq:ideal_sens} is a quasi-polynomial. However, notice that the harmonic zeros that we prescribed through the reference sensitivity \eqref{eq:sens_poly} are placed exactly on the imaginary axis at the desired positions. The required filtration performance for the prescribed zeros is documented in \figref{fig:bode} with the frequency response of the sensitivity \eqref{eq:ideal_sens}. As can be seen, the magnitude features zero values for the frequencies $f_i= 2i \, \SI{}{[Hz]}, i=1..8$ imposed by the design.  The frequency response also reveals the performance of the closed-loop from two robustness aspects, namely, the $H_\infty$-norm and the robustness against frequency variation. Having $\left| S(\mathrm{j}\omega) \right|_\infty < 2$ suggests that the closed-loop has a relatively good robustness against plant/model mismatch. 



In order to demonstrate the robustness of the formed control system, the ideal assumption is dropped by perturbing the system by $G_s(s) = G_m(s) \frac{0.9}{0.05s+1}$. The bottom graph of \figref{fig:spectrum} shows the resulting spectrum for the perturbed case, in which both zeros and poles are now infinitely many due to the fact that the sensitivity corresponds to an infinite-dimensional system of retarded type. Nevertheless, despite steering away from the finite-dimensional case, notice that the placed zeros are not affected by this perturbation and the poles are still in the stable region. As can also be seen, with the growing magnitude of the poles, they tend to match the position of the zeros.  

\subsection{Experimental Results}

\begin{figure}[t]
    \centering
    \includegraphics[width=0.9\columnwidth]{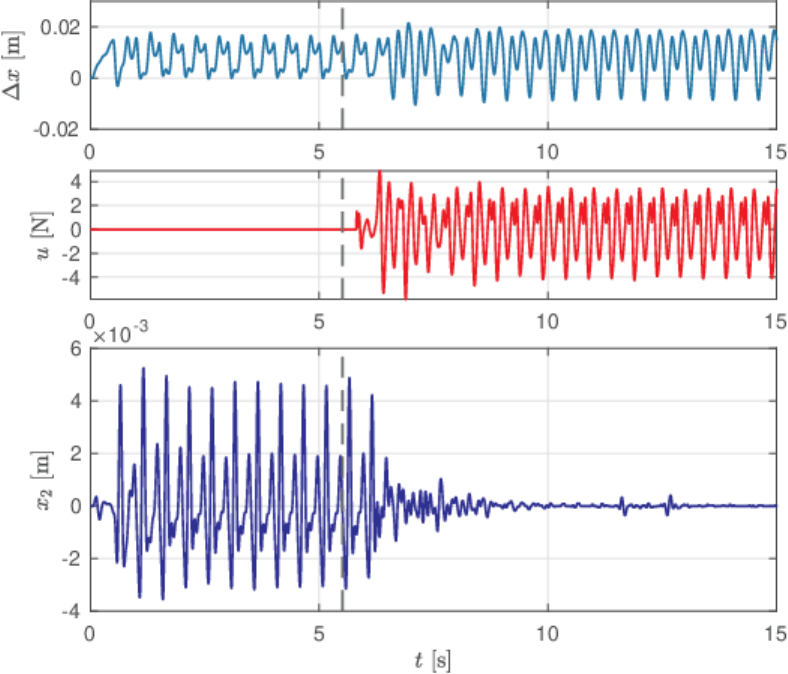}
    \caption{Disturbance rejection performance of the proposed controller. The vertical dashed line indicates the activation moment $t=\SI{5.5}{\second}$ of the controller. \textbf{(Top)} Transient behavior of the resonating structure relative to mass $m_1$ i.e. $\Delta x = x_1-x_a$ \textbf{(Middle)} Control signal u given by the IMC controller. \textbf{(Bottom)} Measured position of mass $m_2$. }
    \label{fig:suppresstion}
\end{figure}

The proposed IMC controller \eqref{IMCfinal1}-\eqref{IMCfinal2} is discretized by a zero-order hold method and implemented in {LabVIEW\textsuperscript{\texttrademark}} on the setup hardware described in Section 3.1. As the results of the experiments are close to being ideal, we omit simulation-based analysis and present the experimental results directly in \figref{fig:suppresstion}. 
In the time range $t\in [0, 5.5]\,\mathrm{s}$ we can observe the passive response of the setup to the $\SI{2}{Hz}$ saw-tooth excitation at the disturbance force $d_F$. As can be seen, it results in a distinct periodic motion at the controlled output $x_2$. A single period of passive motion is shown in \figref{fig:disturbance} accompanied by spectral analysis already discussed above. The controller is turned on at $t=\SI{5.5}{\second}$ and, as can be seen from the middle graph, starts generating the control action after a small delay caused by the controller's own. Once the control action starts to take effect, almost ideal rejection is achieved as clearly visible in the bottom graph. 
The exemplary performance of the proposed controller is demonstrated in video\footnote{\url{https://control.fs.cvut.cz/en/aclab/experiments/imcpz}} of another experiment with a longer time range, where, additionally, the reference tracking ability is demonstrated. 

\section{Conclusion}

As stated by the Internal Model Principle, the design of controllers for tracking/rejection essentially corresponds to a stabilization problem. In fact, based on the proposed filter structure \eqref{eq:ideal_filter}, which can be viewed as a structured Youla-Ku\v{c}era parameter, it was shown that this stabilization problem can be reduced to the stabilization of the chosen signal model. Additionally, forming the controller in the Internal Model Control configuration not only helped the closed-loop to be stable when the system suffers from delays but also revealed that the negative effects of system delay on tracking/rejection can be compensated by an additional delay in the controller. Within this framework, the controller can be designed in a straightforward manner completely analytically for a given stable system without non-minimum phase zeros as in \eqref{eq:G}. The findings were validated both numerically and experimentally. 

Future research directions include extension of the method to stable non-minimum phase systems and to unstable systems. The former shall be solvable within the IMC framework by removing the non-minimum phase zeros from the inverted part of the system dynamics used within the controller and adjusting  
the design formulas and filter structure accordingly. For the unstable systems, an application of an extended scheme arising from Youla-Ku\v{c}era parameterization will be investigated.   

\bibliographystyle{plain}        
\bibliography{References}  

\end{document}